\pdfoutput=1 %
\documentclass[fleqn,10pt]{wlscirep}
\usepackage[utf8]{inputenc}
\usepackage[T1]{fontenc}
\usepackage{subcaption}
\usepackage[framemethod=default]{mdframed}
\usepackage[utf8]{inputenc}
\usepackage{tikz}
\usetikzlibrary{arrows.meta}
\usepackage{graphicx}
\usepackage{geometry}
\usepackage{anyfontsize}
\usepackage{circuitikz}

\title{\textbf{\LARGE{Digital interventions and habit formation in educational technology}}\thanks{ We are grateful to Deepak Agarwal and the team at Stones2Milestones for collaboration on this project. The Golub Capital Social Impact Lab at Stanford Graduate School of Business provided funding for this research.}\\ \vspace{0.2cm} \normalsize }

\author[1]{Keshav Agrawal}
\author[1]{Susan Athey}
\author[2]{Ayush Kanodia}
\author[1]{Emil Palikot}
\affil[1]{Stanford University, Graduate School of Business, Stanford, USA}
\affil[2]{Stanford University, Computer Science, Stanford, USA}

\date{August 2023}

\begin{abstract}
We evaluate a contest-based intervention intended to increase the usage of an educational app that helps children in India learn to read English. The evaluation included approximately 10,000 children, of whom about half were randomly selected to enter a reading contest, whereby those children who ranked sufficiently high on a leaderboard were awarded a set of books. During 12 weeks after the contest, when the treatment group had no additional incentives to use the app, children in the treated group read 75\% more stories than the control group, consistent with the formation of reading habits. Furthermore, post-contest, the treated group abandoned the app 6\% less often than the control group. These results demonstrate that low-cost interventions have the potential to be used to instill reading habits in the digital context.
\end{abstract}

\begin{document}
\flushbottom
\maketitle

\section{Main}

In recent years, digital applications and services developed by the educational technology (EdTech) sector have attracted millions of students \cite{hashim2018teched, coursera2021impact, blanco2022duolingo, boa2023}. These tools offer a variety of potential advantages, including ease of access, adaptability, and cost-efficiency \cite{brasca2022mckinsey}. However, especially when offered directly to students, these applications and services often struggle to keep students engaged and ensure course completion \cite{unesco2023gem}. A significant number of learners face difficulties in establishing consistent usage habits, resulting in high dropout rates \cite{clow2013moocs, jordan2014initial, khalil2014moocs, hone2016exploring, romero2020determining}. Factors contributing to this include dwindling motivation, time management struggles, and the lack of real-time feedback \cite{kizilcec2013deconstructing, badali2022motivation}. 

In contrast, digital entertainment and social media products use a variety of tactics to keep students engaged, often becoming a part of their daily routine \cite{kemp2020, zenith}. The design and engagement strategies employed in these domains have been so effective that their pervasive usage patterns are sometimes likened to addictive behaviors \cite{allcott2022digital, alter2017irresistible, newport2019digital, eyal2019indistractable}. Nonetheless, the successes of these entertainment-oriented products provide valuable insights into effective engagement strategies. The behavioral techniques and design principles they utilize may be adapted for educational settings \cite{huang2021combating}, and some tactics, such as gamification, have been analyzed as critical components of educational applications \cite{pham2022developing}. This paper evaluates the effects of incorporating a contest for students into an EdTech app, a feature that gamifies the app experience, focusing on its influence on student engagement during the contest and subsequent long-term usage patterns.

One approach to enhancing long-term student engagement involves incentivizing the use of the app by new students, thereby instilling learning habits\cite{lopez2022algorithmic}. However, a potential concern with incentives relates to the "crowding-out" effect. This phenomenon, documented in children's psychology \cite{lepper1973undermining,gneezy2011and}, emphasizes the potential for intrinsic learning motivation to be overshadowed or even suppressed by external incentives or rewards. Consequently, a successful incentive-based intervention designed to instill habits needs two components. First, it must be effective during the treatment period, that is, students who are incentivized to engage with the product do, in fact, comply. This can be difficult to achieve, however, as several experimental studies in the education context have evaluated interventions based on financial incentives and have found that they can be ineffective \cite{fryer2011financial, gneezy2000pay, cameron1994reinforcement,bettinger2012paying}. There is also mixed evidence for the effectiveness of behavioral nudges in education \cite{damgaard2018nudging, levitt2016behavioralist,kizilcec2020scaling}. Second, after removing the incentive, treated students need to maintain a higher level of engagement, something that may not occur if the incentive crowds out intrinsic motivation. Whether the two conditions are met may depend on the context of the intervention, including factors such as the experimental setting and the individual characteristics of the experimental subjects \cite{buyalskaya2023can}, and there remains much to learn about the effectiveness of such interventions in the EdTech context. Some suggestive evidence about gamification in educational settings, such as leaderboards in a teacher education course, suggests that gamification can enhance short-term engagement in the classroom, as well as performance\cite{ccakirouglu2017gamifying}, but there is little evidence about scalable interventions in digital education.

This paper seeks to address this gap by providing evidence from a randomized controlled trial of an intervention designed to boost reading on an online educational platform for children in India learning English. Treated students entered a ``100 days reading contest,'' where within the app, a digital ``leaderboard'' displayed a ranking of students based on the number of stories read, and where the students were informed that the top 1,000 students on the leaderboard would win a set of books. First, we show that this intervention was effective during the contest. Treated students read 0.96 (standard error (SE) 0.32) more stories than the control group, 45\% (SE 14\%) higher than the control group mean of 2.12.  Second, during the twelve weeks after the contest, when there were no additional incentives to use the app, these students still read 0.30 (SE 0.14) more stories than the control group students, who, on average, completed 0.40 stories. Finally, many students left the app during the post-contest period; however, treated students were more likely to continue usage, with an estimated increase in 12-week retention of 1.3 percentage points (pp) (SE 0.3 pp) relative to the control group mean of 19\%.  These results are consistent with the hypothesis that the contest intervention reinforced, rather than stifled, students' inherent reading motivation.

\subsection{Empirical framework}

Our study was conducted in the context of an application named \emph{Freadom}, developed by an Indian firm \emph{Stones2Mi\-lestones} (S2M). \emph{Freadom} serves children learning to read in English and, at the time of our study, had approximately 55,000 unique monthly students. Students discover \emph{Freadom} through two main channels: direct consumer advertising and school-based outreach. Our study focuses on students who use \emph{Freadom} through this second channel, whereby S2M forms a partnership with the school and then teachers encourage their students to use the application. School-based students can access the main features free of charge.  

The app's core content consists of short, illustrated stories, sourced from publishers specializing in children's educational content. Upon accessing the app, students are served stories, where stories are selected through a mix of human curation and algorithmic assignment based on a recommendation system \cite{agrawal2022personalized}. After clicking on a story title, a student views a brief 2 to 5-sentence description, an accompanying image, and a "start" button. Students can either start reading or exit the description. Completion of a story prompts an optional set of 3 to 5 multiple-choice questions to assess comprehension. Reading the story and answering the questions typically requires no more than 10 minutes.

\subsection{Design of the intervention and outcomes}

\begin{figure}
    \centering
    \caption{Screenshots of the Freadom App}

    \subcaptionbox{Home Feed Page\label{fig:feed_page}}[0.25\textwidth][c]{%
        \includegraphics[width=0.25\textwidth]{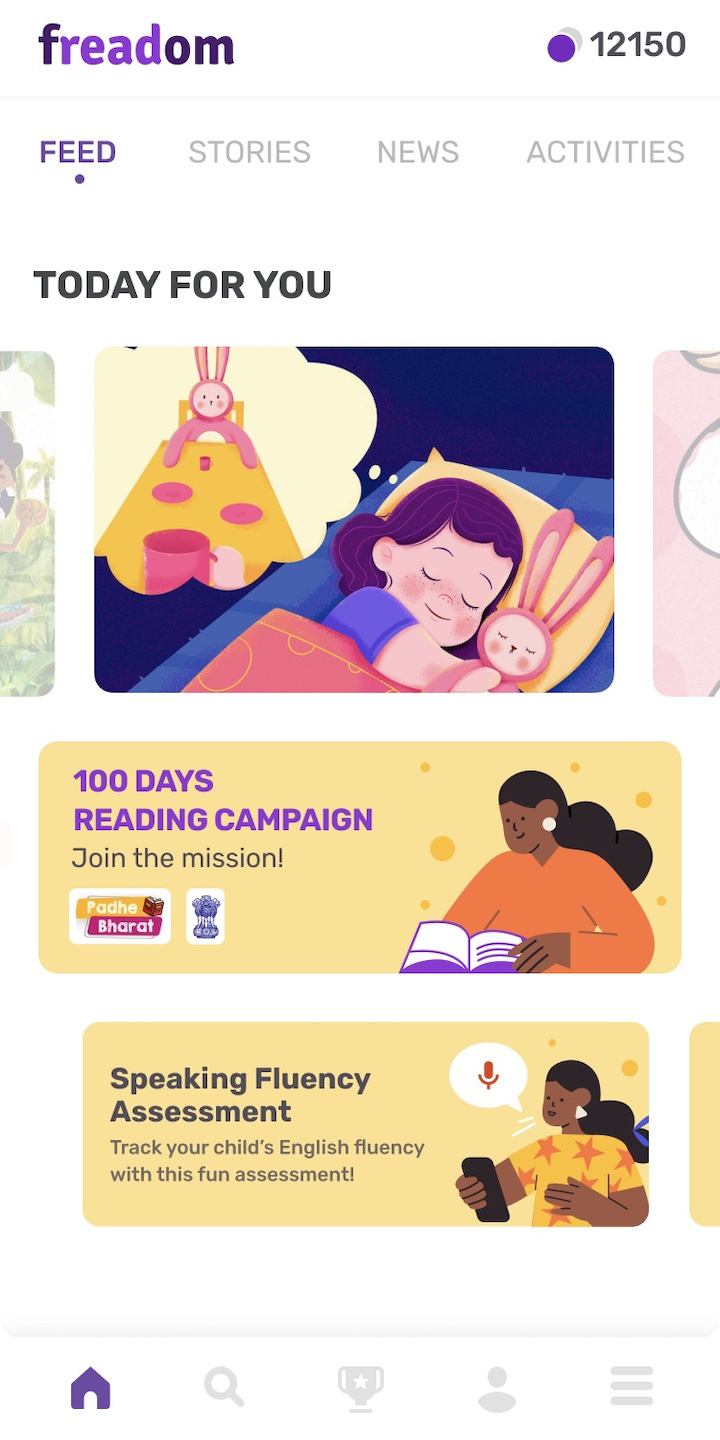}}
    \hfill
    \subcaptionbox{Leaderboard Page\label{fig:leaderboard_sub}}[0.25\textwidth][c]{%
        \includegraphics[width=0.25\textwidth]{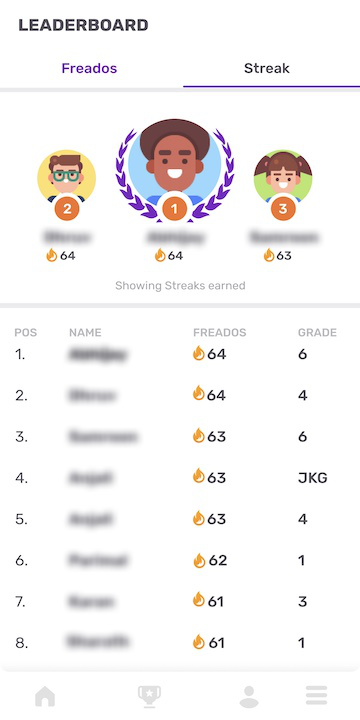}}
    \hfill
    \subcaptionbox{Reading Challenge Promotion\label{fig:reading_stats}}[0.25\textwidth][c]{%
        \includegraphics[width=0.25\textwidth]{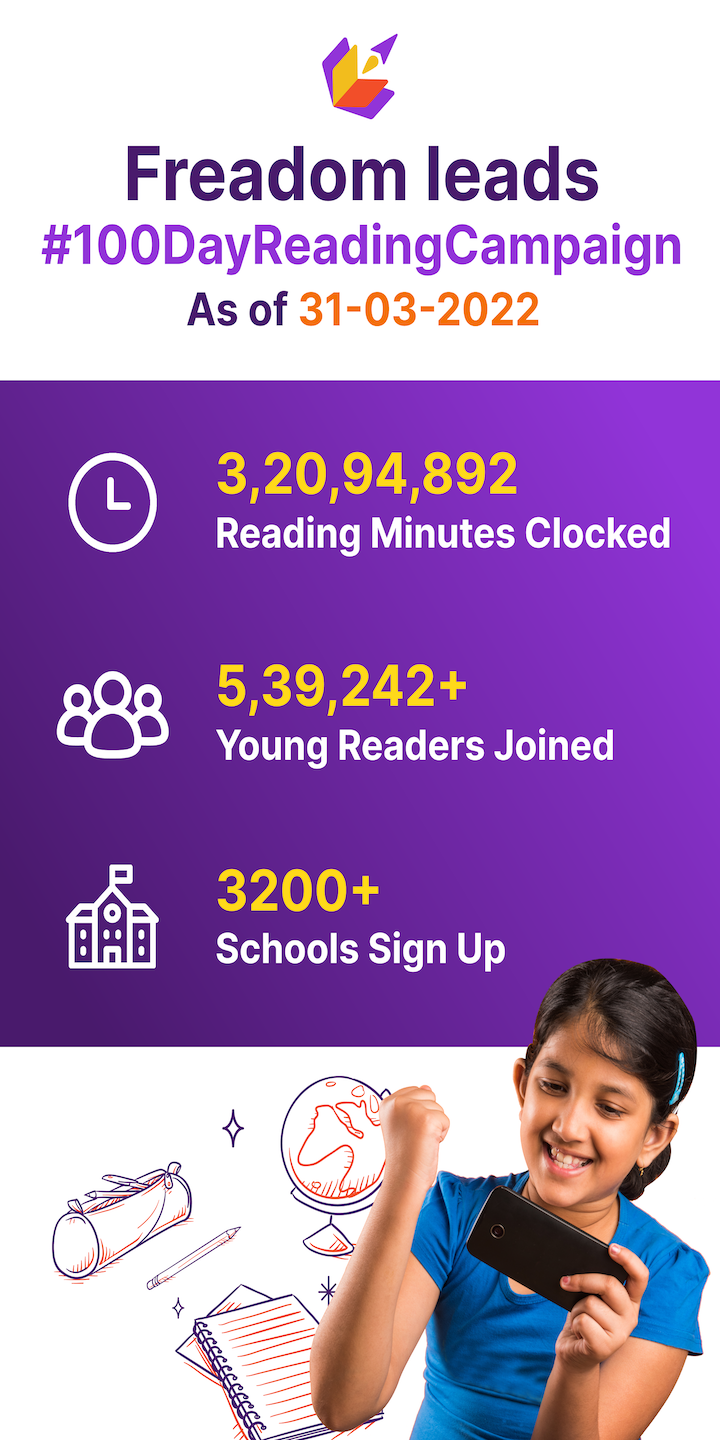}}

    \caption*{\footnotesize{\textit{Note: The three images are screenshots from the Freadom app. Figure \ref{fig:feed_page} is the landing page of the app. Figure \ref{fig:leaderboard_sub} is the leaderboard that showcases the top students in the Reading Challenge. Finally, Figure \ref{fig:reading_stats} is one of the promotion banners created by \emph{Stones2Milestones} to promote the campaign.}}}

    \label{fig:leaderboard}
\end{figure}

During our intervention, students in the treatment group were ranked on a leaderboard based on the number of stories they completed, as shown in Figure \ref{fig:leaderboard}. Students ranked in the top 1,000 at the end of the contest were awarded a set of six books delivered to their homes. The control group had access to the standard leaderboard offered in \emph{Freadom}, which ranked students based on various in-app activities and was less prominently displayed in the app. The control group did not have rewards for ranking higher on this leaderboard.

\begin{figure}
\centering
    \caption{Experiment timeline}
\resizebox{1\textwidth}{!}{%
\begin{circuitikz}
\tikzstyle{every node}=[font=\large]
\draw [ color={rgb,255:red,32; green,48; blue,172}, , line width=1pt](33.75,9) to[short] (42.25,9);
\node [font=\large, color={rgb,255:red,32; green,48; blue,172}] at (37.75,9.5) {Post-treatment period};
\draw [ color={rgb,255:red,181; green,203; blue,11}, , line width=1pt](22.75,12.75) to[short] (33.75,12.75);
\draw [ color={rgb,255:red,181; green,203; blue,11}, , line width=1pt](23.75,11.5) to[short] (33.75,11.5);
\draw [ color={rgb,255:red,181; green,203; blue,11}, , line width=1pt](25,10.25) to[short] (33.75,10.25);
\draw [ color={rgb,255:red,181; green,203; blue,11}, , line width=1pt](26.25,9) to[short] (33.75,9);
\draw [ color={rgb,255:red,181; green,203; blue,11}, , line width=1pt](27.5,7.75) to[short] (33.75,7.75);
\draw [ color={rgb,255:red,181; green,203; blue,11}, , line width=1pt](28.75,6.5) to[short] (33.75,6.5);
\draw [ color={rgb,255:red,181; green,203; blue,11}, , line width=1pt](30,5.25) to[short] (33.75,5.25);
\node [font=\large, color={rgb,255:red,181; green,203; blue,11}] at (28.5,13.25) {Treatment period};
\node [font=\large, color={rgb,255:red,13; green,13; blue,13}] at (34,3.25) {End of treatment: 05/31/2022};
\node [font=\large, color={rgb,255:red,13; green,13; blue,13}] at (42,3.25) {End of data collection: 11/30/2022};
\draw [ color={rgb,255:red,193; green,21; blue,21} , fill={rgb,255:red,193; green,22; blue,22}] (22.75,12.75) circle (0.25cm);
\draw [ color={rgb,255:red,193; green,21; blue,21} , fill={rgb,255:red,193; green,22; blue,22}] (23.75,11.5) circle (0.25cm);
\draw [ color={rgb,255:red,193; green,21; blue,21} , fill={rgb,255:red,193; green,22; blue,22}] (25,10.25) circle (0.25cm);
\draw [ color={rgb,255:red,193; green,21; blue,21} , fill={rgb,255:red,193; green,22; blue,22}] (26.25,9) circle (0.25cm);
\draw [ color={rgb,255:red,193; green,21; blue,21} , fill={rgb,255:red,193; green,22; blue,22}] (27.5,7.75) circle (0.25cm);
\draw [ color={rgb,255:red,193; green,21; blue,21} , fill={rgb,255:red,193; green,22; blue,22}] (28.75,6.5) circle (0.25cm);
\draw [ color={rgb,255:red,193; green,21; blue,21} , fill={rgb,255:red,193; green,22; blue,22}] (30,5.25) circle (0.25cm);
\node [font=\normalsize, color={rgb,255:red,193; green,21; blue,21}] at (23,12.25) {Randomization 1: 02/10/2022};
\node [font=\normalsize, color={rgb,255:red,193; green,21; blue,21}] at (23.5,10.75) {Randomization 2: 02/18/2022};
\node [font=\normalsize, color={rgb,255:red,193; green,21; blue,21}] at (25,9.75) {Randomization 3: 02/25/2022};
\node [font=\normalsize, color={rgb,255:red,193; green,21; blue,21}] at (26.25,8.5) {Randomization 4: 03/05/2022};
\node [font=\normalsize, color={rgb,255:red,193; green,21; blue,21}] at (27.5,7.25) {Randomization 5: 03/21/2022};
\node [font=\normalsize, color={rgb,255:red,193; green,21; blue,21}] at (28.75,6) {Randomization 6: 04/04/2022};
\node [font=\normalsize, color={rgb,255:red,193; green,21; blue,21}] at (30.25,4.5) {Randomization 7: 04/16/2022};
\draw [ color={rgb,255:red,193; green,21; blue,21}, , line width=1pt](20,12.75) to[short] (22.5,12.75);
\draw [ color={rgb,255:red,193; green,21; blue,21}, , line width=1pt](22.5,11.5) to[short] (23.75,11.5);
\draw [ color={rgb,255:red,193; green,21; blue,21}, , line width=1pt](23.75,10.25) to[short] (25,10.25);
\draw [ color={rgb,255:red,193; green,21; blue,21}, , line width=1pt](25,9) to[short] (26.25,9);
\draw [ color={rgb,255:red,193; green,21; blue,21}, , line width=1pt](26.25,7.75) to[short] (27.5,7.75);
\draw [ color={rgb,255:red,193; green,21; blue,21}, , line width=1pt](27.5,6.5) to[short] (28.5,6.5);
\draw [ color={rgb,255:red,193; green,21; blue,21}, , line width=1pt](28.75,5.25) to[short] (29.75,5.25);
\node [font=\large, color={rgb,255:red,193; green,21; blue,21}] at (21.5,13.5) {Pre-treatment period};
\draw [ color={rgb,255:red,5; green,5; blue,5} , fill={rgb,255:red,8; green,7; blue,7}, line width=1pt ] (20,12.75) circle (0.25cm);
\draw [ color={rgb,255:red,5; green,5; blue,5} , fill={rgb,255:red,8; green,7; blue,7}, line width=1pt ] (22.75,11.5) circle (0.25cm);
\draw [ color={rgb,255:red,5; green,5; blue,5} , fill={rgb,255:red,8; green,7; blue,7}, line width=1pt ] (23.75,10.25) circle (0.25cm);
\draw [ color={rgb,255:red,5; green,5; blue,5} , fill={rgb,255:red,8; green,7; blue,7}, line width=1pt ] (25,9) circle (0.25cm);
\draw [ color={rgb,255:red,5; green,5; blue,5} , fill={rgb,255:red,8; green,7; blue,7}, line width=1pt ] (26.25,7.75) circle (0.25cm);
\draw [ color={rgb,255:red,5; green,5; blue,5} , fill={rgb,255:red,8; green,7; blue,7}, line width=1pt ] (27.5,6.5) circle (0.25cm);
\draw [ color={rgb,255:red,5; green,5; blue,5} , fill={rgb,255:red,8; green,7; blue,7}, line width=1pt ] (28.75,5.25) circle (0.25cm);
\node [font=\normalsize, color={rgb,255:red,5; green,5; blue,5}] at (19.5,11.5) {Batch 2: 30 schools \&  2082 students};
\node [font=\normalsize, color={rgb,255:red,5; green,5; blue,5}] at (16.75,12.75) {Batch 1: 25 schools \& 2938 students};
\node [font=\normalsize, color={rgb,255:red,5; green,5; blue,5}] at (20.5,10.25) {Batch 3: 27 schools \&  1073 students};
\node [font=\normalsize, color={rgb,255:red,5; green,5; blue,5}] at (21.5,9) {Batch 4: 14 schools \&  850 students};
\node [font=\normalsize, color={rgb,255:red,5; green,5; blue,5}] at (22.75,7.75) {Batch 5: 14 schools \&  1149 students};
\node [font=\normalsize, color={rgb,255:red,5; green,5; blue,5}] at (24.25,6.5) {Batch 6: 16 schools \&  781 students};
\node [font=\normalsize, color={rgb,255:red,5; green,5; blue,5}] at (25.25,5) {Batch 7: 12 schools \&  464 students};
\node [font=\large, color={rgb,255:red,5; green,5; blue,5}] at (19.75,3) {Start of the experiment: 02/01/2022};
\draw [ color={rgb,255:red,5; green,5; blue,5}, line width=1pt, ->, >=Stealth] (17.75,4) .. controls (31,4) and (31,4) .. (44,4);
\draw [ color={rgb,255:red,5; green,5; blue,5}, line width=1pt, dashed] (33.75,12.75) .. controls (33.75,8.5) and (33.75,8.5) .. (33.75,4.25);
\draw [ color={rgb,255:red,5; green,5; blue,5}, line width=1pt, dashed] (42.5,12.75) .. controls (42.5,8.5) and (42.5,8.5) .. (42.5,4.25);
\draw [ color={rgb,255:red,5; green,5; blue,5}, line width=1pt, dashed] (22.75,12.75) .. controls (22.75,12) and (22.75,12) .. (22.75,11.5);
\draw [ color={rgb,255:red,5; green,5; blue,5}, line width=1pt, dashed] (23.75,11.5) .. controls (23.75,10.75) and (23.75,10.75) .. (23.75,10.25);
\draw [ color={rgb,255:red,5; green,5; blue,5}, line width=1pt, dashed] (25,10) .. controls (25,9.5) and (25,9.5) .. (25,9);
\draw [ color={rgb,255:red,5; green,5; blue,5}, line width=1pt, dashed] (26.25,9) .. controls (26.25,8.5) and (26.25,8.5) .. (26.25,8);
\draw [ color={rgb,255:red,5; green,5; blue,5}, line width=1pt, dashed] (27.5,7.5) .. controls (27.5,7) and (27.5,7) .. (27.5,6.5);
\draw [ color={rgb,255:red,5; green,5; blue,5}, line width=1pt, dashed] (28.75,6.5) .. controls (28.75,6) and (28.75,6) .. (28.75,5.5);
\draw [ color={rgb,255:red,5; green,5; blue,5}, , line width=1pt](20,4.25) to[short] (20,3.75);
\draw [ color={rgb,255:red,5; green,5; blue,5}, , line width=1pt](22.5,4.25) to[short] (22.5,3.75);
\draw [ color={rgb,255:red,5; green,5; blue,5}, , line width=1pt](23.75,4.25) to[short] (23.75,3.75);
\draw [ color={rgb,255:red,5; green,5; blue,5}, , line width=1pt](25,4.25) to[short] (25,3.75);
\draw [ color={rgb,255:red,5; green,5; blue,5}, , line width=1pt](26.25,4.25) to[short] (26.25,3.75);
\draw [ color={rgb,255:red,5; green,5; blue,5}, , line width=1pt](27.5,4.25) to[short] (27.5,3.75);
\draw [ color={rgb,255:red,5; green,5; blue,5}, , line width=1pt](28.75,4.25) to[short] (28.75,3.75);
\draw [ color={rgb,255:red,5; green,5; blue,5}, , line width=1pt](30,4) to[short] (30,4.25);
\draw [ color={rgb,255:red,5; green,5; blue,5}, , line width=1pt](30,4.25) to[short] (30,3.75);
\end{circuitikz}
}%
    \caption*{\footnotesize{\textit{Note: In total, 138 schools were randomized, which resulted in 9337 students. Randomization was stratified based on the school size, size of grades, and city.}}}\label{design_graph}
\end{figure}
We evaluated this contest using a randomized experiment, where the treatment was assigned at the level of a school so that all students in a given school were in the same experimental group.  A school became eligible for the experiment if its leadership decided to collaborate with S2M and encouraged students to use the \emph{Freadom} app during our experiment's period, which ran from February 1, 2022, to May 31, 2022. The timing of the enrollment for the experiment was aligned with India's academic calendar, where the academic year ends in late February to mid-March, and a new academic year starts in late March to mid-April. Thus, for many students, the start of the contest occurred just before their school break. We included 138 schools in the experiment. In the contest, we considered only students in Grades 1 to 6, which resulted in 9337 students.

Figure \ref{design_graph} presents the timeline of the experiment. The time period in which schools entered the experiment is partitioned into seven consecutive time intervals, and we refer to the interval in which a school entered the experiment as its batch. After a school was enrolled in the experiment, teachers began informing the students about the app.  For each batch, all of the schools in the batch (that is, those that entered during the associated time interval) were randomized into the treatment or the control group. On average, 8 days elapsed between a student's registration in the app and the experimental group assignment (for details, see Section \ref{days_appendix}). Subsequently, students from the schools in the treatment group were introduced to the contest and its associated leaderboard through in-app notifications, direct alerts, and WhatsApp messages to their parents. There were no differences in other app features between the experimental groups. However, the notifications and reminders might have made the treated students more aware of the app in general rather than specifically of the contest. Students who registered for the app after the randomization of their school occurred were not included in our experimental sample and did not participate in the contest.

Given the relatively small number of schools, the experimental design included stratification.  In particular, the sample was stratified based on school size, number of students per grade, and the city, and randomization occurred within each group. 

 For each student, we refer to the period between the start of their batch and the randomization as the \emph{pre-treatment} phase. Figure \ref{design_graph} shows the phase's start and end dates for each batch.  After randomization into one of the experimental groups, \emph{treatment period} starts. Due to the differences in the time of randomization, this phase had different duration for students in different batches. In this phase, treated students received notifications about the contest and competed for a high leaderboard ranking. Data from this phase gauges the intervention's engagement impact. To assess the lasting effects of the intervention, we use the data collected during the \emph{post-treatment} period, extending up to 24 weeks after the end of the contest. This phase has the same duration for all batches; it started on May 31, 2022, and lasted until November 31, 2022.

Our primary outcomes are \emph{Total Stories Completed} and \emph{Total Story Engagement}. \emph{Total Stories Completed} represents the total number of stories completed in a considered period, while \emph{Total Story Engagement} is defined by weighing student-story interactions: 0.3 for viewing a story's description but not starting it, 0.5 for starting a story but not finishing it, and 1.0 for completing a story. The \emph{Total Story Engagement} represents the sum of all \emph{Story Engagement} scores over a specific period.  We also consider \emph{Retention}, which is defined only in the \emph{post-treatment} period; for students that have viewed at least one story in the \emph{post-treatment}, \emph{Retention} takes the value of one in weeks from the end of the contest until the week in which they viewed their last story, and zero later on, and for students that did not view any stories in \emph{post-treatment} period takes the value of zero in all weeks. A supplementary outcome is the number of visits to the leaderboard page on the app -- \emph{Total Leaderboard Visits}.

Table \ref{tab:sum_stats} details summary statistics of characteristics of schools in the treatment and control groups. We find no significant differences between the treatment and control groups. Table \ref{tab:diffmeans} shows that the pre-treatment differences across treatment and control groups in utilization outcomes are not statistically distinguishable from zero.  Specifically, the utilization of the app as measured by \emph{Total Stories Completed}, \emph{Total Story Engagement}, and \emph{Total Leaderboard Visits} is similar across the two experimental groups.

\begin{table}
     \caption{Summary statistics of the main variables}
     \label{tab:sum_stats}
      \centering
          \resizebox{0.9\textwidth}{!}{%
      \begin{tabular}{ l c c *{8}{cc}}
        \toprule
        \toprule
      Variables & \multicolumn{4}{c}{Treated group} & \multicolumn{4}{c}{Control group} \\ 
      \cmidrule(l){2-5} \cmidrule(l){6-9}
         & Mean & St. Dev. & Min & Max &Mean & St. Dev. & Min & Max\\ 
      \midrule
     Grade Level & 3.853 & 0.699 & 2.333 & 5.076 & 3.853 & 0.762 & 1.785 & 5.400 \\ 
Students enrolled per school & 65.23 & 95.09 & 1 & 542 & 70.56 & 83.21 & 4 & 384 \\
School Fees (in INR) & 51659.43 & 37545.09 & 10500 & 216000 & 76479.11 & 116499.98 & 10900 & 709400 \\
Large City & 0.533 & 0.502 & 0 & 1 & 0.508 & 0.504 & 0 & 1 \\
Number Days before Assignment & 8.201 & 9.242 & 0 & 68.285 & 7.628 & 4.424 & 1.086 & 23.25 \\

     \bottomrule
     \bottomrule
       \end{tabular}
       }
       \caption*{\footnotesize{\textit{Note: Number of units (schools) in the treated group is 75 and in the control group is 63, which results in 4887 students in the treatment group and 4440 in the control group. Summary statistics of per school average covariate values. The variables include the Grade Level of students, categorized into six levels; the number of students per school; school fees; and a binary indicator binary indicator of whether a school is located in large city or not based on the stratification of cities by S2M; and the number of days between student activation and experimental group assignment.}}}
     \end{table}
\section{Results}

\begin{center}
\begin{figure}
    \caption{Student engagement over time across treatment and control}
    \includegraphics[width=0.9\textwidth]{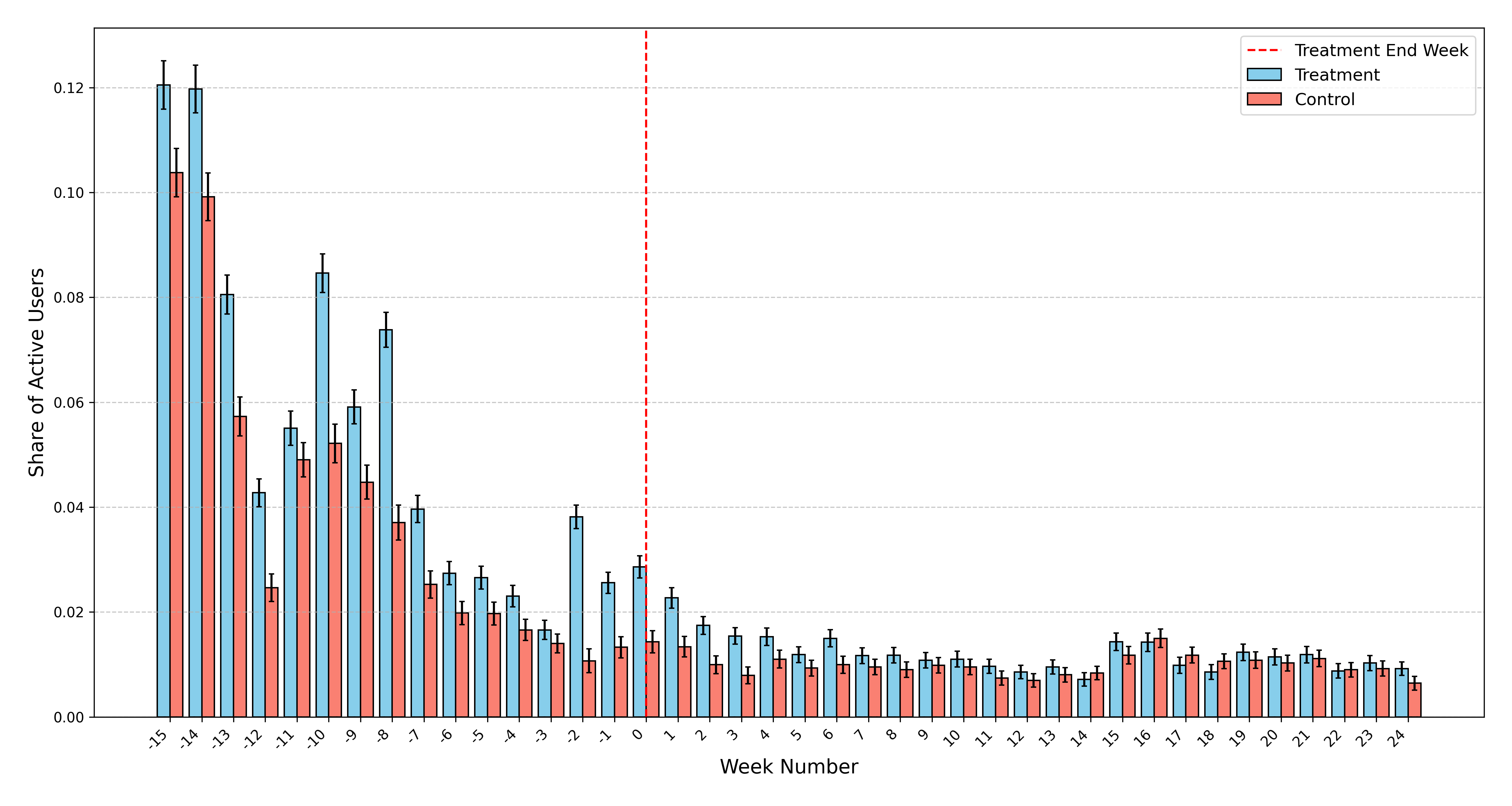}
    \caption*{\footnotesize{\textit{Note: y-axis shows the weekly fraction of students that completed at least one story across treatment and control groups. X-axis shows weeks relative to the end of the experiment. The treatment group is in blue, and the control group is in orange. Whiskers show confidence intervals. }}}
    \label{fig:story-completion-trend}
\end{figure}
\end{center}

Figure \ref{fig:story-completion-trend} shows the weekly shares of students in the treatment and control groups who completed at least one story. Weeks are ordered such that week 0 is the first week after the contest. We note a clear disparity in story completion rates between the treatment and control groups during the contest. This difference is especially marked immediately after the beginning of the experiment when students were informed of the potential rewards and right before the end of the contest, when some treated students increased their engagement with the app, possibly aiming to improve their leaderboard rankings and secure the rewards. Following the termination of the treatment phase and subsequent prize distribution, a sustained higher engagement level from the treatment group is discernible. Despite the natural drop-off, engagement levels remained higher for the treatment group until the end of the considered period 24 weeks post-contest.

\subsection{Average treatment effects}

Table~\ref{tab:diffmeans} presents the estimates of the average treatment effects, estimated using the difference-in-means estimator for \emph{Total Stories Completed}, \emph{Total Story Engagement}, and \emph{Total Leaderboard Visits} and using the Cox Proportional Hazard model\cite{cox1972regression} for \emph{Retention}.

First, we find no statistically significant differences between experimental groups in pre-treatment app usage, which is expected given random assignment of learners to experimental groups. Second, the treatment group had almost twice as many visits to the leaderboard as the control group during the contest, consistent with the intervention's design, where leaderboard rankings influenced prize attainment.  Further, we estimate a large and statistically significant increase in app usage during the experimental phase in the treatment group: treated students completed on average an additional 0.96 story (SE 0.32) and increased their \emph{Total Story Engagement} by 1.09 (SE 0.37), corresponding to increases of 45\% and 44\% over the respective baseline averages. Four weeks after the contest ended, relative to the control group, the treatment group had 0.18 (SE 0.07) more completed stories and an increase in engagement of 0.20 (SE 0.08). Students in the treatment group were 9.8\% (or 2.2pp with a standard error of 0.008) more inclined to continue using the app across the four weeks after the end of the contest. The differences persist through the \emph{post-treatment} period.

\begin{table}[ht!]
\caption{Estimates of the Average Treatment Effects}
\label{tab:diffmeans}
\centering
 \resizebox{0.65\textwidth}{!}{%
\begin{tabular}{lccccccc}
\toprule
\toprule
\textbf{Outcome Description} & \textbf{\small{Mean Treatment}} & \textbf{\small{Mean Control}} & \textbf{ATE} & \textbf{SE} & \textbf{CI lb} & \textbf{CI ub}\\ \hline 
      \addlinespace 
\multicolumn{7}{l}{\emph{Pre-Treatment Period}} \\ 
\addlinespace 
Total Stories Completed & 1.878 & 2.112 & -0.233 & 0.198 & -0.621 & 0.155 \\
Total Story Engagement & 2.397 & 2.720 & -0.323 & 0.235 & -0.783 & 0.138 \\
Total Leaderboard Visits & 1.897 & 1.849 & 0.048 & 0.196 & -0.336 & 0.433 \\
\hline 
\addlinespace 
\multicolumn{7}{l}{\emph{Treatment Period}} \\ 
\addlinespace 
Total Stories Completed & 3.084 & 2.123 & 0.961 & 0.323 & 0.328 & 1.594 \\
Total Story Engagement & 3.550 & 2.460 & 1.091 & 0.372 & 0.362 & 1.819 \\
Total Leaderboard Visits & 1.893 & 0.984 & 0.909 & 0.263 & 0.393 & 1.425 \\
\hline 
\addlinespace 
\multicolumn{7}{l}{\emph{4 Weeks Post-Treatment Period}} \\ 
\addlinespace 
Total Stories Completed & 0.324 & 0.146 & 0.178 & 0.073 & 0.036 & 0.320 \\
Total Story Engagement & 0.374 & 0.170 & 0.203 & 0.084 & 0.039 & 0.368 \\
Total Leaderboard Visits & 0.211 & 0.0585 & 0.152 & 0.066 & 0.023 & 0.282 \\
\hline 
\addlinespace 
\multicolumn{7}{l}{\emph{12 Weeks Post-Treatment Period}} \\ 
\addlinespace 
Total Stories Completed & 0.695 & 0.395 & 0.301 & 0.139 & 0.028 & 0.573 \\
Total Story Engagement & 0.814 & 0.467 & 0.347 & 0.161 & 0.031 & 0.662 \\
Total Leaderboard Visits & 0.458 & 0.181 & 0.277 & 0.112 & 0.057 & 0.498 \\
\hline 
\addlinespace

\multicolumn{7}{l}{\emph{Student Retention}} \\ \addlinespace 
4 Weeks Post-Treatment & 0.246 & 0.225 & 0.022 & 0.008 & 0.006 & 0.037\\
12 Weeks Post-Treatment & 0.203 & 0.190 & 0.012 & 0.003 & 0.006 & 0.018\\
24 Weeks Post-Treatment & 0.152 & 0.143 & 0.009 & 0.001 & 0.007 & 0.011\\

\bottomrule
\bottomrule
\end{tabular}
}
\caption*{\footnotesize{\textit{Note: Estimates of the average treatment effect using difference-in-means estimator for Total Story Completed, Total Story Engagement, and Total Leaderboard Visits, and from Cox Proportional Hazard model for \emph{Retention}. In the top panel, outcomes in the pre-treatment period; in the second panel, outcomes during the contest; and in the third panel, outcomes after the end of the treatment. In the bottom panel, the estimates for Retention using the Cox model with controls. The number of students in the treatment group is 4887, and in the control group is 4440. For \emph{Retention} outcomes, there are 4 observations per student in 4 week Post-Treatment, 12 in 12 weeks Post-Treatment, and 24 in 24 weeks Post-treatment. Standard errors are clustered at the school level for Total Story Completed, Total Story Engagement, and Total Leaderboard Visits and at the student level for \emph{Retention}. 
}}}
\end{table}

\subsection{Heterogenous treatment effects}

In Table \ref{tab:heterogeneity_results}, we present estimates of the average treatment effect per subgroup based on four characteristics: \emph{heavy student} -- which takes the value of 1 when the student has over median \emph{Total Story Engagement} in the \emph{pre-treatment} period, \emph{Large City} -- for students coming from large cities (defined by \emph{Stones2Milestones}), and \emph{Older Students} -- students that are over median age (age is deduced from school grade). The outcome metrics under consideration are post-experimental \emph{Total Story Engagement} and \emph{Retention}. For the former, we employ the difference-in-means estimator, while for the latter the Cox Proportional Hazard model.

Regarding the impact on \emph{Total Story Engagement}, our results indicate that the treatment effects during the post-contest phase were more pronounced for students attending less expensive schools, younger students, and those who exhibited higher engagement levels before the experiment. We identify similar patterns of treatment effect heterogeneity with \emph{Retention} as the outcome; additionally, we find that participants from smaller cities had higher treatment effects than those from larger ones.

\begin{table}
        \caption{Conditional average treatment effects}
         \label{tab:heterogeneity_results}
      \centering
          \resizebox{0.75\textwidth}{!}{%
\begin{tabular}{ l *{5}{c} }
    \toprule
    \toprule
  & \multicolumn{2}{c}{Variable value is 1} & \multicolumn{2}{c}{Variable value is 0} & \\ 
  \cmidrule(l){2-3} \cmidrule(l){4-5}
     Variable & Baseline & CATE & Baseline & CATE & Diff. Subgroups \\ 
     \midrule
  \addlinespace
    \multicolumn{5}{l}{\emph{Total Story Engagement (4 weeks Post-Treatment)}} \\ \addlinespace 
    High usage & 0.346 (0.044) & 0.436 (0.093) & 0.084 (0.011) & 0.115 (0.029) & 0.321 (0.098) \\
    Old student & 0.122 (0.015) & 0.164 (0.039) & 0.234 (0.033) & 0.268 (0.065) & -0.104 (0.076) \\
    Large city & 0.197 (0.026) & 0.215 (0.053) & 0.138 (0.018) & 0.172 (0.041) & 0.043 (0.067) \\
    Expensive school & 0.149 (0.026) & 0.109 (0.046) & 0.186 (0.021) & 0.276 (0.052) & -0.167 (0.069) \\
    \midrule 
  \addlinespace 
    \multicolumn{5}{l}{\emph{Retention at 12 weeks}} \\ \addlinespace   
    High usage & 24.900 (0.028) & 1.620 (0.002) & 15.040 (0.019) & 0.977 (0.001) & 0.640 (0.002) \\
    Old student & 16.020 (0.021) & 1.040 (0.001) & 23.110 (0.028) & 1.500 (0.002) & -0.460 (0.002) \\
    Large city & 20.580 (0.021) & 1.340 (0.001) & 16.800 (0.030) & 1.090 (0.002) & 0.245 (0.002) \\
    Expensive school & 16.660 (0.029) & 1.080 (0.002) & 20.740 (0.021) & 1.350 (0.001) & -0.265 (0.002) \\
    \bottomrule
    \bottomrule
\end{tabular}
       }
       \caption*{\footnotesize{\textit{Note: Conditional average treatment effects per subgroup. In the top panel, the outcome variable is post-experiment Story Engagement. In the bottom panel, Retention of students post-experiment in percentage. Students are considered to be active until the week in which they completed their last story. Heavy students are students with over median Story Engagement in the pre-experimental period; Large city students are students from larger cities (as defined by \emph{Stones2Milestones}), and Older students are students above median age. An expensive school is a school with fees higher the median. Columns one and three show outcomes in control groups. Columns two and four are conditional average treatment effects estimates from the difference-in-means estimator in the case of Story Engagement and the Cox proportional hazard model for Retention. The last column is the difference between CATEs. Standard errors are at the student level for Retention.
       }}}
     \end{table}

\subsection{Additional analyses suggestive of mechanisms}

\subsubsection{Change in student app behavior}
Entering a contest in which students compete against each other, coupled with a prominent ranking system, might change the way students interact with the app. Post-contest engagement might not solely be contingent on the volume of interaction during the contest, but also the nature of such interactions. Now we delineate three potential types of app engagement during the contest: (i) \emph{late-heavy}:  we identify students who exhibited a spike in their interaction towards the contest's end. Students whose percentage of \emph{Total Story Engagement} in the last month exceeded the median of the control group are designated as \emph{late-heavy}. Two primary reasons can account for this behavior. Some students might have intensified their usage with the aim of clinching the prize, which might not lead to sustained post-contest app usage. Conversely, this behavior might reflect students' extended app engagement following an initial exploration phase; (ii) \emph{competitive}: students with \emph{Total Leaderboard Visits} surpassing the median of the control group during the contest are termed \emph{competitive} students. Given that our study introduced participants to the leaderboard, it is plausible that students who became familiar with this feature might change app engagement after the end of the contest; (iii) \emph{steady}: students who completed at least one story monthly throughout the contest are labeled as \emph{steady} students. Regular reading could indicate a consistent usage habit, potentially enhancing retention after the end of the contest.
\begin{table}[ht!]
\centering
\caption{Usage patterns analysis}%
\label{tab:mediation}
          \resizebox{0.55\textwidth}{!}{%
\begin{tabular}{lccccc}
\toprule
\toprule
\textbf{Treatment} & \textbf{Pattern} &  \textbf{Share of Students} & \textbf{\emph{Retention}} & \textbf{\emph{Total Story Engagement}} \\ \hline

\midrule

\multicolumn{5}{l}{\emph{Competitive Student}} \\ \addlinespace   
1 & 1 & 0.641 (0.009)  & 0.209 (0.011) & 1.955 (0.198) \\ 
1 & 0 & 0.484 (0.006) & 0.069 (0.004) & 0.308 (0.036) \\ 
0 & 1 & 0.359 (0.009) & 0.221 (0.014) & 1.438 (0.171) \\
0 & 0 & 0.516 (0.006) & 0.086 (0.005) & 0.241 (0.027) \\ 
\midrule
\multicolumn{5}{l}{\emph{Late Heavy Student}} \\ \addlinespace   
1 & 1 & 0.634 (0.015) & 0.372 (0.019) & 4.397 (0.441) \\ 
1 & 0 & 0.511 (0.005) & 0.073 (0.004) & 0.279 (0.031) \\ 
0 & 1 & 0.366 (0.015) & 0.391 (0.026) & 3.146 (0.406) \\ 
0 & 0 & 0.489 (0.005) & 0.086 (0.004) & 0.226 (0.020) \\ 
\midrule
\multicolumn{5}{l}{\emph{Steady Student}} \\ \addlinespace   
1 & 1 & 0.539 (0.008) & 0.154 (0.009) & 1.546 (0.168) \\ 
1 & 0 & 0.515 (0.006) & 0.088 (0.005) & 0.397 (0.040) \\
0 & 1 & 0.461 (0.008) & 0.154 (0.009) & 0.839 (0.102) \\ 
0 & 0 & 0.485 (0.006) & 0.089 (0.005) & 0.275 (0.028) \\

    \bottomrule
     \bottomrule
\end{tabular}
}
\caption*{\footnotesize{\textit{Note: The top panel shows the results using \emph{late-heavy} indicator, central panel the \emph{competitive student} indicator, and  the bottom panel \emph{steady student} variable. The first column shows the value of the usage pattern indicator; the second column is the treatment indicator. The third column shows the shares of the experimental subgroup (treatment or control) that exhibit the usage pattern during the contest. The last two columns show retention and engagement outcomes in the post-treatment period in the group defined by the value of the in-app behavior indicator and treatment indicator, together with standard errors.
}}}
\end{table}

Table \ref{tab:mediation} compares the shares of students that were \emph{late-heavy}, \emph{competitive}, and \emph{steady} across the treatment groups. The share of \emph{late-heavy} students increased due to treatment from 37\% to 63\% (difference of 26 pp), the share of \emph{competitive} students increased from 36\% to 64\% (difference of 28 pp), and the share of \emph{steady} students increases from 46\% to 54\% (8pp difference). Second, the \emph{late-heavy}, \emph{competitive}, and \emph{steady} students have higher outcomes than their respective counterparts. For example, \emph{late-heavy} students are much more likely to be retained for twelve weeks: 37\% and 39\% of \emph{late-heavy} students are retained in the treatment and control groups, respectively, compared to 7.3\% and 8.6\% of students who are not \emph{late-heavy}; this type of students also have higher engagement. These results suggest that the intervention changed the way students interact with the app by making them more likely to exhibit patterns of app usage typically associated with heavy engagement and increased retention.

\subsubsection{Consumption of difficult content}
While we do not have direct measures of learning, we observe the difficulty level of completed stories. All stories in \emph{S2M} catalog are labeled by publishers as appropriate for a specific grade. If a student completes a story above their grade, we consider the story as difficult for that student -- stories at the student's grade or below we label as not difficult.  Here, we focus on the number of difficult stories completed by each student and the share of difficult stories in all completed stories. The share of difficult stories is a ratio of difficult stories to all stories completed by that student. If a student did not complete any stories, we assign zero. When the number or the share of difficult stories increases, we consider this to be indicative of learning. We analyze the impact of treatment on these outcomes during the experiment and in the post-experiment period. 

New \emph{Freadom} students do not receive personalized recommendations; instead, they all observe the same ranked list of stories from which they select the content they want to read. Consequently, the increase in the number of difficult stories or their share is not driven by the algorithmic selection but by students' choices. Furthermore, the S2M catalog includes thousands of stories, making it unlikely that a student has exhausted all stories in their grade, thereby necessitating a shift to more difficult ones.

Using a Poisson regression, we estimate that treated students read 0.33 (SE 0.13) more difficult stories during the contest. After the end of the contest, the treatment group completed, on average, 0.54 (SE 0.23) more difficult stories. The estimates of treatment impact on the share of difficult stories in the total number of completed stories are positive but statistically insignificant; using a difference-in-means estimator, we find an increase of the share of difficult stories by 0.043 (SE 0.037) during the contest and by 0.021 (SE 0.016) after the end of it. 

The difficulty level of completed stories is an imperfect proxy for learning; for example, the difficulty might be correlated with some other characteristics or imperfectly measured. Nevertheless, the finding that students select and complete more difficult stories is suggestive of learning.

We also find that treated students who completed more difficult stories in the \emph{pre-treatment} period have larger treatment effects with \emph{Total Story Engagement} as the outcome. Adjusting for the \emph{pre-treatment} total number of story interactions, the number of difficult stories completed, and the interaction of treatment and pre-contest engagement, we estimate that an additional difficult story completed pre-contest increases the treatment effect by 0.46 (SE 0.10) during the contest and 0.04 (SE 0.02) in the period after it. This suggests that learners who had higher levels of comprehension, to begin with, engaged more with the contest and were more likely to form a learning habit.

\subsubsection{Decomposition of the treatment effect into higher Retention and higher utilization}\label{decompo_subsection}
\begin{figure}
    \caption{Retention and engagement increase decomposition}\label{fig:ret_eng}
    \centering
    \includegraphics[width=0.6\textwidth]{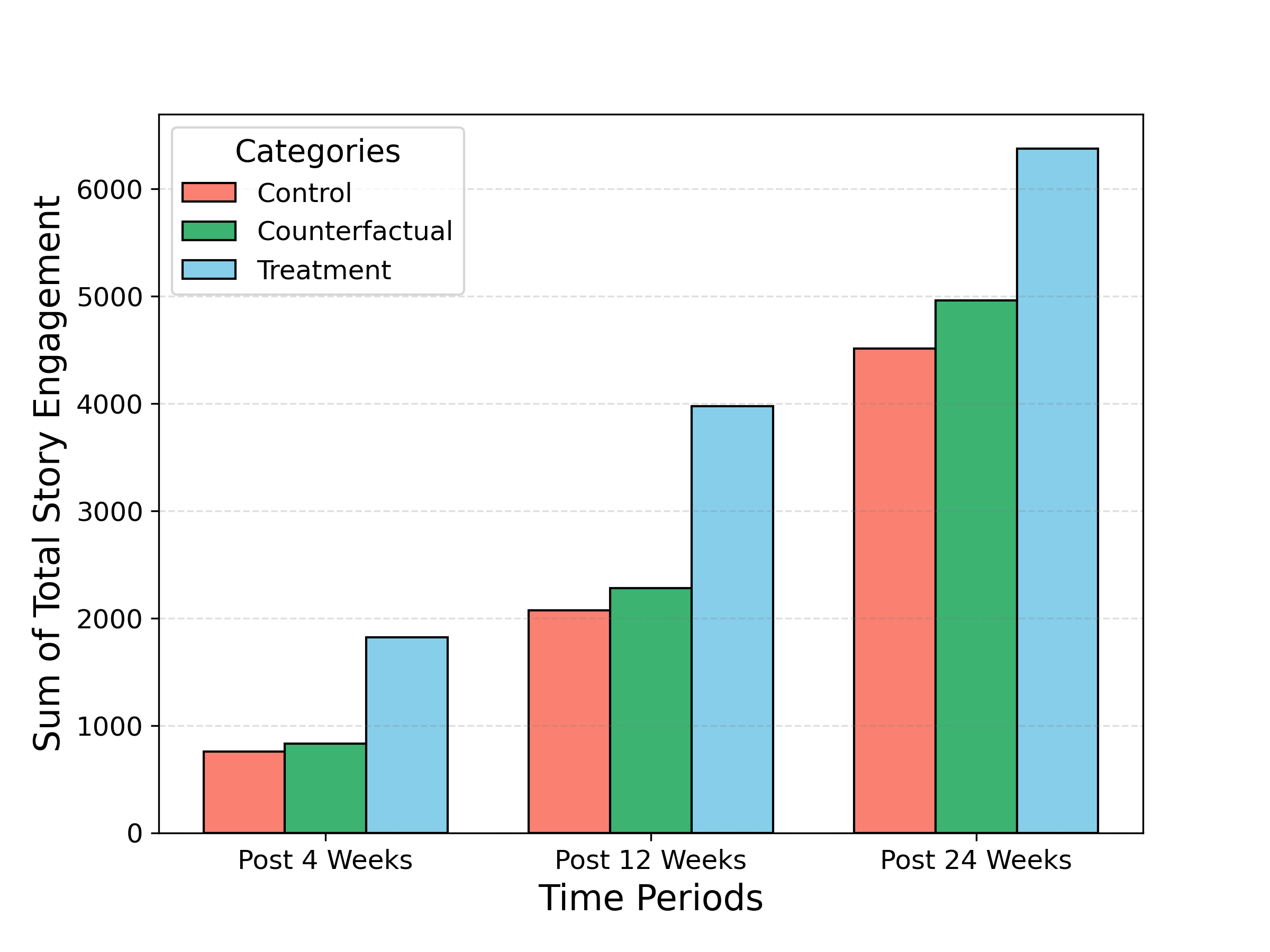}
    \caption*{\footnotesize{\textit{Note: Decomposition of the impact of increased Retention and higher Total Story Engagement of active students. Orange bars are sums of the Total Story Engagement of the control group. Green bars -- the counts of active students in the treatment group multiplied by the average Total Story Engagement in the control group. The blue bars show the sums of the Total Story engagement in the treatment group. Sums of Total Story Engagement are cumulative over time.
}}}
\end{figure}

The change in the engagement across the two experiment groups can be decomposed into higher \emph{Retention}, which results in utilization of the app by students that would have otherwise churned, and higher \emph{Total Story Engagement} of students that would have stayed on the app in the absence of treatment. We present this decomposition in Figure \ref{fig:ret_eng}.

We use the sum of \emph{Total Story Engagement} across all students in an experimental group as the outcome of interest: we consider the first 4 weeks after the end of the contest, the first 12 weeks, and 24 weeks. First, we compute this outcome for the control group (orange bars in Figure \ref{fig:ret_eng}). Two factors change between the periods: a rise in \emph{Total Story Engagement} as students engage with more stories over time and activation of students that were inactive in previous weeks (e.g., a student could have been inactive in the first 4 weeks after treatment, but returned to the app in the week after that). Second, we repeat the same computation for the treatment group (blue bars in Figure \ref{fig:ret_eng}). Differences between the treated and control groups are due to the differences in the number of active students in each period and their average \emph{Total Story Engagement}. Finally, to decompose these two factors, we carry out a counterfactual computation, where for each unit in the treatment group, we keep the student's \emph{Retention} but replace the student's actual \emph{Total Story Engagement} in the considered period with the average \emph{Total Story Engagement} in the control group. Thus, if the student has been retained, we assign the average outcome in the control group, and if the student has not been retained, we keep a zero level of \emph{Total Story Engagement} (green bars in Figure \ref{fig:ret_eng}). This way, we isolate the incremental benefits stemming purely from higher \emph{Retention}.

The comparison of the control group and the counterfactual outcomes shows that while enhanced \emph{Retention} certainly amplifies the sum of \emph{Total Story Engagement} in the treatment group, its contribution is modest. In contrast, comparing counterfactual with treatment, which isolates the impact attributable solely to per-student change in \emph{Total Story Engagement}, shows a high difference. We conclude that the gain of the average \emph{Total Story Engagement} rather than higher \emph{Retention} is the primary driver of the treatment effect.

\section{Discussion}
The rise of Educational Technology (EdTech) has ushered in a plethora of learning platforms catering to diverse audiences, with mobile apps and MOOCs alone having accrued 440 million users by the end of 2022. However, challenges persist. Many EdTech products suffer from attrition, as students do not succeed in forming consistent usage habits. This study seeks to understand whether interventions commonly used in entertainment-oriented digital apps can bolster student engagement in educational apps and foster long-term usage habits. We provide encouraging results in this context by introducing a contest-style intervention into a reading app geared towards children in India learning English and carrying out a randomized experiment.

During the contest, treated students displayed a marked increase in reading activity, outpacing the control group by 45\% (SE 14\%) more completed stories. Twelve weeks after the contest, these students persisted in their heightened engagement, reading more stories and remaining more engaged than their counterparts in the control group, even without a leaderboard or any other additional incentives. Notably, the intervention also impacts the typically high attrition rates seen in EdTech, with treated students showcasing a significant uptick in 12-week retention rates by 1.3 pp (SE 0.3 pp) relative to the control group retention rate of 19\%. This lends credence to the idea that the contest paradigm did not merely overlay a superficial layer of engagement but strengthened the intrinsic motivation to read.

Despite these promising results, a salient limitation of our study is the lack of insight into the actual progress in learning. While engagement metrics such as story completion rates (including the difficulty level of completed stories) and app retention are important and indicative of learning, an important goal of any educational tool is to foster genuine, impactful learning. Thus, while our findings shed light on strategies to heighten engagement in EdTech platforms, further research is needed to discern whether such enhanced engagement translates to tangible educational outcomes and progress in learning.

\section{Methods}
\subsection{IRB}

The experiment and the partnership with \emph{Stones2Milestones} was approved by the Stanford IRB eProtocol \#64003.

\subsection{Treatment assignment}\label{days_appendix}

\begin{figure}
    \caption{Histograms of the number of days of app use before randomization}\label{fig:his_days}
    \centering
    \includegraphics[width=0.85\textwidth]{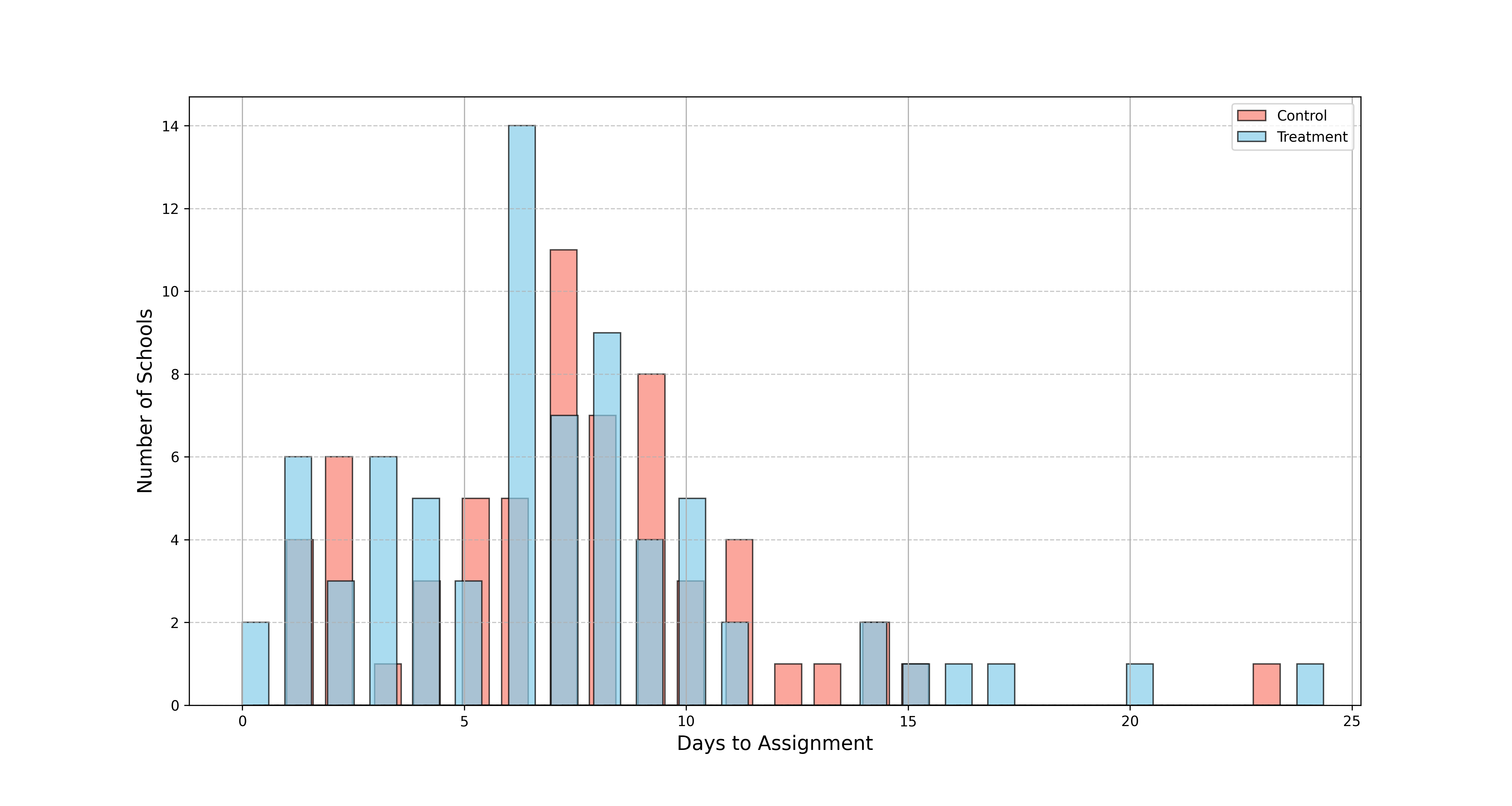}
    \caption*{\footnotesize{\textit{Note: Histograms of the average per school number of days between students' app activation and of treatment assignments.
}}}
\end{figure}

Schools were assigned to experimental groups in seven batches on the following dates: February 10, 18, 25, March 5, 21, and April 4 and 16 (refer to Figure \ref{design_graph} for a visual representation). For each randomization, we incorporated all schools that S2M had enrolled since the previous batch (except for the initial batch, which included schools enrolled between February 1 and February 10). This resulted in variations in the duration of app usage by students prior to treatment assignment. Figure \ref{fig:his_days} presents histograms illustrating the distribution of the number of days students were active in the app before being assigned to the treatment or control group.

We find that the difference between treatment and control in the average of the number of days prior to randomization is statistically insignificant (the difference is 0.572 with a standard error of 1.272). Furthermore, the Kolmogorov-Smirnov test has a p-value of 0.316.

\subsection{Data collection}

In this research, we use data collected by \emph{Stones2Milestones} (S2M) on their \emph{Freadom} app. To preserve student privacy, all data was fully anonymized by \emph{Stones2Milestones}. 
The data encompassed two integral components: app navigation information and student and story characteristics. The app navigation data contains indicators and timestamps of students' interaction with all app elements. From this, we learn which students launched the app and when and which sections of the app they visited. Of particular interest is data on student-story interactions. We observe which stories students selected to view and also whether they started and completed these stories. Key outcome variables are constructed based on this information. In addition, data on student characteristics comes from app registration and contains the school name, grade, and location. Based on schools' names, \emph{S2M} provided information on school fees. Finally, we use information on the stories' characteristics and their difficulty level.

\subsection{Data processing}

\paragraph{Construction of zeroes:}
In the experiment, the treated students received app notifications and communications through external channels, particularly WhatsApp messages. As a consequence, we consider all students in the treated group as treated and opening the app or not as an outcome. In all analyses, students who did not launch the app during the studied period are assigned zero values on all outcome variables. In contrast, dropping students who did not launch the app would bias our results due to the exclusion of students who were not motivated by the possibility of winning the book set enough to launch the app.

\paragraph{Outlier adjustment:} Even though our data comes directly from the \emph{Freadom} student tracking system and represents the most accurate, available information, there are occasional instrumentation errors. These errors might result in a lack of tracking information available for certain sections of the app for certain days or misattribution of app navigation. Misattribution might result in spurious high app engagement. To account for this, we remove the students who have greater than 99.9\% of total completed stories in a period. Our results are robust to changing the trimming cutoff to 99\% and 95\%.

\subsection{Poisson and Cox Proportional Hazard Estimators}

The Poisson regression model is utilized to model the number of difficult stories students consume. The model is described by Equation \ref{poisson}.
\begin{equation}\label{poisson}
\log(E[Y|X, W]) = \alpha*W +  X\beta,
\end{equation}

where $Y$ is the number of difficult stories, $W$ is the treatment indicator, and $X$ are other covariates. We are primarily interested in the coefficient $\alpha$. A significant assumption of the Poisson model is the equidispersion assumption, which states that the conditional mean is equal to the conditional variance. 

We use the Cox Proportional Hazard model\cite{cox1972regression} in all analyses of \emph{Retention}. Formally, the model is described in Equation \ref{cox},
\begin{equation}\label{cox}
h(t|X,W) = h_0(t) \exp(\alpha W + X\beta),
\end{equation}
Where $h(t|X,W)$ denotes the hazard function given covariates  $X$ and treatment indicator $W$, $h_0(t)$ is the baseline hazard function at time $t$, and $\exp(\alpha W + X\beta)$ denotes the relative risk associated with the covariates. The model's primary assumption is the proportionality of hazards, meaning the effects of the predictors are multiplicatively consistent over time.

\bibliography{refs}

\end{document}